\documentclass[%
 reprint,
 amsmath,amssymb,
 aps,
]{revtex4-2}

\usepackage{graphicx}
\usepackage{dcolumn}
\usepackage{bm}

\usepackage{xcolor}

\usepackage{pdfpages}

\makeatletter
\AtBeginDocument{\let\LS@rot\@undefined}
\makeatother

\begin{document}


\title{Stabilization of the skyrmion in a hybrid magnetic--superconducting nanostucture}

\author{Julia Kharlan$^{1,2}$, Mateusz Zelent$^{1}$, Konstantin Guslienko$^{3}$, Vladimir O. Golub$^{2}$, Jarosław W. Kłos$^{1}$}
\email{yuliia.kharlan@amu.edu.pl}
\affiliation{
$^{1}$ISQI, Faculty of Physics, Adam Mickiewicz University, Poznań, Poland\\
$^{2}$V.G. Barykhtar Institute NASU of Ukraine, Kyiv, Ukraine
}%
\affiliation{$^3$Depto. Polimeros y Materiales Avanzados: Fisica, Quimica y Tecnologia, Universidad del País Vasco, UPV/EHU, 20018 San Sebastian, Spain\\
EHU Quantum Center, University of the Basque Country, UPV/EHU, 48940 Leioa, Spain\\
IKERBASQUE, the Basque Foundation for Science, 48013 Bilbao, Spain}

\date{\today}

\begin{abstract} 
Stabilization of skyrmions in a magnetic material without the Dzyaloshinskii–Moriya exchange interaction requires using of an inhomogeneous magnetic field, such as a demagnetization field or an Oersted field. To control and tune  the local magnetic field in magnetic material we propose to exploit superconducting nanorings. The field stabilizes the skyrmion through the presence of persistent current induced by the pulses of external field.  We analyze the conditions for the stabilization of Néel skyrmion  in ferromagnetic layers with out-of-plane anisotropy, as a function of the nanoring size and induced   superconducting current. We show that the superconducting current should exceed a critical value for the skyrmion to become stable. The paper presents consistent results from both analytical and micromagnetic calculations for Co and Ga:YIG thin magnetic films.
\end{abstract}

\maketitle


\section{Introduction \label{sec:Intro}}
In recent decades, inhomogeneous magnetic textures such as domain walls, skyrmions, vortices, etc., attracted much attention because of their potential applications in spintronics and magnonics \cite{dieny_2020,YU20211}. Domain walls and skyrmions can be used to store and transmit the information in racetrack memory systems \cite{fert2017,parkin_2008}. Moreover, the information can be processed on the neuromorphic platforms based on magnetic vortices or skyrmions \cite{kurenkov_2019,song_2020}.  

Conventionally, magnetic skyrmions \cite{fert2017,Li_2023} used to be stabilized by the Dzyaloshinskii-Moriya interaction (DMI) \cite{Valle_2015,aranda2018,guslienko2019}, which imposes rather strong constraints on magnetic materials to be used. To overcome this problem it has been proposed to stabilize the skyrmions by demagnetizing field \cite{Zelent_2025, Kong_2024, montoya2017}. 
Alternatively, it was shown that the Bloch skyrmions can be stabilized in artificial crystals based upon a combination of a out-of-plane magnetized film and nanopatterned arrays of nanodisks \cite{sun2013} or in submicron dots with moderate out-of-plane magnetic anisotropy \cite{guslienko2015,zelent_2017}. Besides, it was demonstrated the stabilization of the Neel-like skyrmions by means
of a strong stray dipolar field generated by a patterned hard magnetic layer placed near the soft magnetic film \cite{Navas2019}. However, all these methods, which play with material parameters, state of interfaces or the sizes of the nanostructures, do not allow  creating and destroying the skyrmions on-demand, making their application limited. To gain such functionality,an external bias is to be used. However, the local control of nanometer-sized skyrmions, e.g., by spin polarized current from STM tip \cite{Romming_2013} is challenging. The simpler approach is to operate with dipolarly stabilized skyrmions of larger sizes and to apply local  magnetic field produced by the electric current.\nobreak This idea was exploited to steer domains walls \cite{alejos_2017}, control magnetization configuration of ferromagnetic (FM) rings \cite{yang_2011}, and induce hopfions in a FM torus \cite{castillo_2021}. 

The skyrmions can be used to design neuromorphic systems\cite{Yokouchi_2022}. Recently, Shan et al. \cite{shan_2024} proposed an on-chip skyrmion artificial synaptic device, where the radii of skyrmions in flat circular nanodots exhibiting interfacial DMI were influenced by the pulses of Oersted field generated by the current flowing in $\Omega$-shaped wires, while the output signal of each synapse was read out from magnetic tunnel junction with a free layer being a manipulated skyrmion \cite{Li_2022}. The authors reasonably point out that one of the problems in this spintronic system is the local heating generated by the currents. Here, we propose an alternative approach to stabilize Néel skyrmions in a continuous ferromagnetic (FM) layer without DMI, using persistent currents induced in superconducting (SC) rings \cite{clem_2011}.
Such on-demand control brings closer the realization of a magnonic neuromorphic system \cite{papp_2021}, where the magnetic skyrmions serve as weights and interconnections of the neural network, while spin waves handle signal routing and nonlinear activation.

The paper has a following structure. The theoretical models developed to calculate the stray magnetic field produced by the SC ring and to determine the radius of a skyrmion stabilized by this field are presented in Sec. II. The  procedure for stabilizing a skyrmion calculated using the AMUMax3 \cite{MathieuMoalicAMUmax} micromagnetic solver is also briefly described at the end of this section. The following Sec. III presents and discusses the outcomes of the theoretical and numerical studies investigating the influence of the SC ring geometry, the magnitude of the SC current, and the material parameters of the FM layer concerning skyrmion stabilization. The studies are summarized with Conclusions. Additional details are provided in the Supplemental Materials, where we show how the dimensions and current of the SC ring affect the skyrmion size, and discuss how the critical current required for the skyrmion stabilization depends on material parameters.

\section{Model \label{sec:Model}}
\begin{figure}[!b]
\includegraphics[width=1.0\columnwidth]{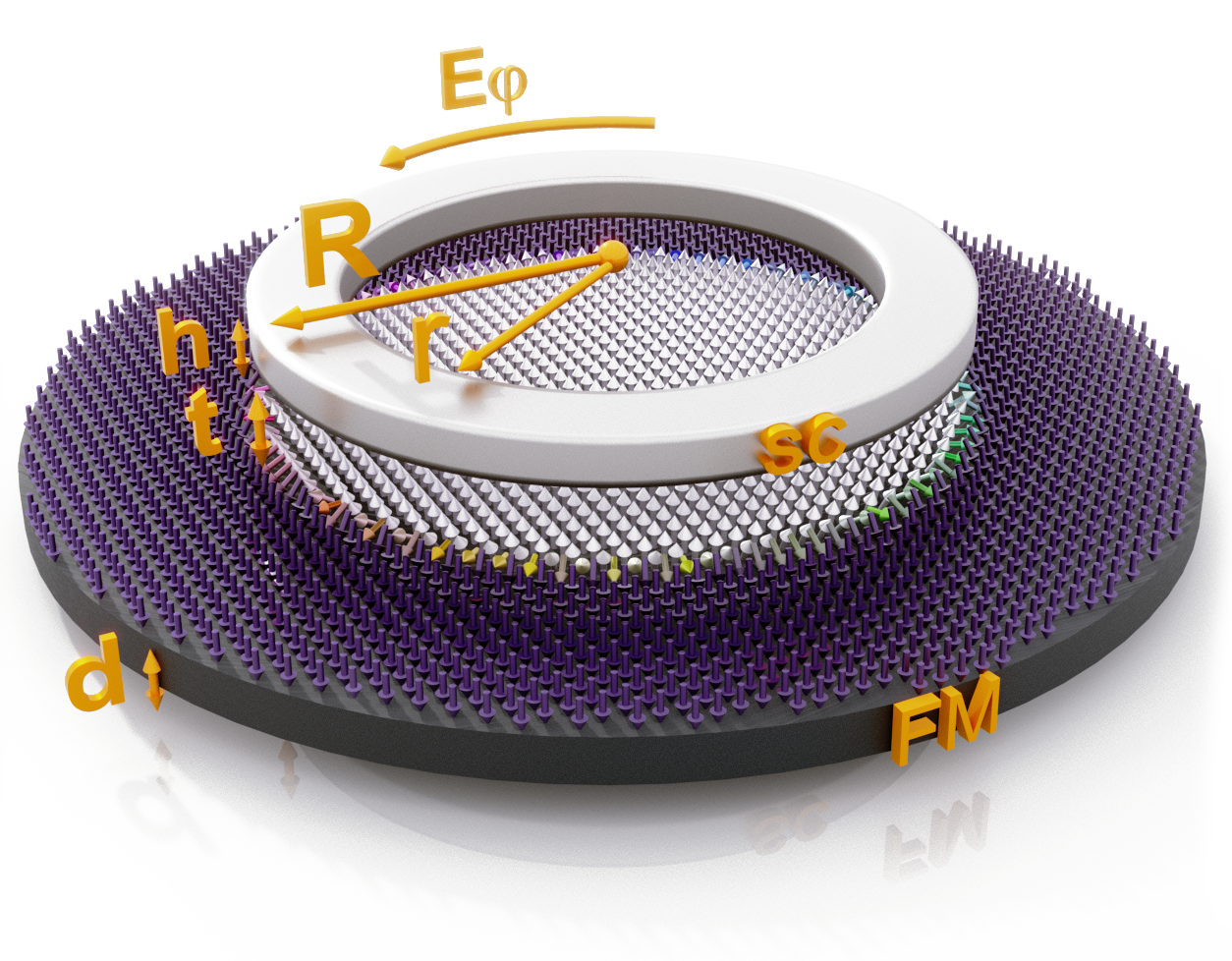}
\caption{
The magnetic stray field of the superconducting (SC) ring stabilizes the magnetization texture in the ferromagnetic (FM) layer with out-of-plane magnetic anisotropy. Eddy currents in the SC ring of external/inner radii $R$, $r$ and height $h$, circulate in a one direction across the entire cross-section due to the electric field component $E_\varphi$. The SC ring is separated from the FM layer by a gap $t$. The thickness $d$ of the FM layer is small compared to the spatial variations in magnetization. The illustrated magnetization texture represents a skyrmion stabilized in gallium-doped YIG—see Fig.~\ref{fig:Skyrmion everything}.
} 
\label{fig:system}
\end{figure}

We consider a single superconducting ring with a thin ferromagnetic layer underneath (see Fig.~\ref{fig:system}). However, our results are valid for an array of SC rings if the distance between them is large enough to exclude inter-ring interactions. The SC ring and the FM layer are electrically isolated from each other. We assume that the SC ring is in the Meissner state and the FM layer is characterized by an effective out-of-plane magnetic anisotropy. To simplify the calculations, we consider a magnetic layer of finite in-plane size, i.e., a circular disk with radius $R_d$, which is larger than the sizes of the SC ring: $R_d > R > r$, where $R$ and $r$ denote the outer and inner radii of the ring, respectively. The SC ring has a rectangular cross section: $(R-r)\times h$, where $h$ is its height, 
and is separated from the FM layer by the gap of the width $t$.
The thickness of the FM layer $d$ is small. Therefore, both magnetic field generated by the SC ring and the FM layer magnetization can be considered as constants over the thickness of the FM layer. Taking into account the cylindrical symmetry of the system, we use the cylindrical coordinates with the radius vector $\mathbf{r}=(\rho, \phi, z)$, where  $z$ is out-of-plane axis.

The SC current can be induced in a superconducting ring using various schemes~\cite{Brandt_2004}, resulting in different spatial distributions of the current density. We consider the case in which a unidirectional flow of SC current is formed throughout the entire cross-section of the ring in the absence of an external magnetic field. In this state, the only source of the static magnetic field is the persistent current, which produces the magnetic field of opposite directions inside and outside the SC ring, thereby enhancing the stabilization of the skyrmion. To reach such a state, we consider the application of an electric field pulse with non-zero curl in the plane of the ring, $\mathbf{E}(t) = E(t)\mathbf{e}_{\varphi}$. We assume a rectangular pulse shape, i.e., the electric field is constant and non-zero during a finite time interval $t_0$.
This effect is clearly demonstrated by the first London equation \cite{Tinkham_2004, Brandt} 
 \begin{equation}
    \mathbf{E}=-\frac{\lambda^{2}}{\mu_0}\partial_t \mathbf{j}\label{eq:London1}
\end{equation}
where the electric field should be turned on only for a short time to change the value of SC current density $\mathbf{j}$. 
The rate of this change depends on the London penetration depth $\lambda$. The symbol $\mu_0$ denotes the permeability of vacuum. The same state can be reached when the SC ring is pooled out of the region of the static magnetic field, having been previously cooled down in the presence of this field.

The currents in the SC ring create a non-uniform distribution of the stray magnetic field with radial symmetry in the FM layer $\mathbf{B}_{\rm sc}(\rho)=B_\rho(\rho)\mathbf{e}_\rho+B_z(\rho) \mathbf{e}_z$, see Fig.~\ref{fig:SC field}. We consider the FM layer (thin disk) with an effective perpendicular magnetic anisotropy (PMA) constant $K=K_u-\mu_0 
M_{\rm s}^2>0$, where the magnetocrystalline ($K_u$) anisotropy overcomes the shape anisotropy of the layer.

The magnetic stray field generated by the SC ring, $\mathbf{B}_{\rm sc}$, stabilizes the Néel-type skyrmion in the FM layer. In principle, one should also take into account the effect of the magnetic field generated by the skyrmion’s magnetization texture on the SC currents, and consequently, on the field produced by the SC ring. However, it can be shown that  this feedback is weak -- the field produced only by the domain wall of the skyrmion is at least an order of magnitude smaller than the stray field of the SC ring. Therefore, we consider only the influence of the SC ring on the FM layer, neglecting the mutual interaction between the two subsystems. Our study was carried out in two consecutive steps. First, we semi-analytically calculate the magnetic stray field generated by the SC ring. Then, we investigate, both analytically and numerically, the stabilization of the Néel-type skyrmion in a thin FM disk placed in the magnetic field produced by the SC ring.

\subsection{The static magnetic field produced by superconducting ring\label{sec:SC_theory}}
To find the magnetic field distribution created by the SC current induced by the electric field, we have modified semi-analytical approach developed in Ref.~\onlinecite{Brandt}. Application of the electric field $\mathbf{E}$ along circumferential direction in the SC ring in Meissner state induces a current of density $\mathbf{j}=j\mathbf{e}_{\varphi}$. 
According to the Ampere equation the current density is connected with  magnetic field $\mathbf{B}_{\rm sc}$ and vector magnetic potential $\mathbf{A}_{\rm sc}$:
\begin{equation}
    \mu_0\mathbf{j} = \nabla\times\mathbf{B}_{\rm sc}=\nabla\times\nabla\times\mathbf{A}_{\rm sc}=-\Delta\mathbf{A}_{\rm sc}\label{eq:amper},
\end{equation}
where the Coulomb gauge was used for the vector potential. Eq.~(\ref{eq:amper}) has a form of Poisson equation for vector potential $\mathbf{A}_{\rm sc}=A_{\rm sc}\mathbf{e}_\varphi$:
\begin{equation}
    \Delta A_{\rm sc}=-\mu_0j+\frac{1}{\rho^2}A_{\rm sc}.\label{eq:poisson}
\end{equation}
The total vector potential $\mathbf{A}= \mathbf{A}_{\rm sc}+\mathbf{A}_{E}$  also includes the contribution $\mathbf{A}_{E}$ originating from the rectangular pulse of applied electric field $\mathbf{E}(t)=E(t)\mathbf{e}_{\varphi}$  \cite{Brandt}:
\begin{equation}
\begin{split}
    &A_{E}=-\int_{0}^{t_0}E dt,
\label{eq:electric}
\end{split}
\end{equation}
where $t_0$ and $E$ are the duration and intensity of the pulse, respectively.
The total vector potential is related to the SC current by the formula \cite{Tinkham_2004}:
\begin{equation}
\begin{split}
    &A=A_{\rm sc}+A_{E}=-\mu_0\lambda^2 j.\label{eq:london}
\end{split}
\end{equation}
derived from the London equation (\ref{eq:London1}), and the relation: $\mathbf{E}=-\partial_t \mathbf{A}$. Combining Eqs.~(\ref{eq:london}),  (\ref{eq:electric}) and (\ref{eq:poisson}), allows to formulate the following integral equation for the density of SC current:
\begin{equation}
\begin{split}
    \lambda^2 j(\mathbf{r})&=\int_{V}  d^{3}\mathbf{r}' Q(\mathbf{r},\mathbf{r'})\left(j(\mathbf{r}')+\frac{\lambda^2}{\rho'^2}\big(j(\mathbf{r}')-j_0\big)\right)\\&+\!\lambda^2j_0, \label{eq:curr_int}
\end{split}
\end{equation}
where $V$ is volume of the SC ring, $j_0=Et_0/(\mu_0\lambda^2)$ and the function $Q=-\tfrac{1}{4\pi|\mathbf{r}-\mathbf{r'}|}$ is the integral kernel for the Laplace operator. 
In the cylindrical coordinate system, it can be expressed in $k$-space using Fourier expansion inside the SC ring \cite{morse_methods_1953}: 
\begin{equation}
\begin{split}
    \frac{1}{|\mathbf{r}-\mathbf{r'}|} = \sum_{n=-\infty}^{+\infty}\int_{0}^{+\infty} dk & \,J_n(k\rho) J_n(k\rho') \\& \times e^{-k|z-z'|} e^{in(\varphi-\varphi')}, 
    \end{split}\label{eq:expansion}
\end{equation}
with $J_n$ being Bessel function of the first kind of n$^{\rm th}$ order. 
Using (\ref{eq:expansion}) and taking into account that $j$ does not depend on $\varphi$-coordinate and weakly depends on $z$--coordinate, Eq.~(\ref{eq:curr_int}) can be simplified to the  form:
\begin{equation}
\begin{split}
    \lambda^2 j(\rho) &=\int_{r}^{R} d\rho' 
    \;\tilde{Q}(\rho,\rho')j(\rho')\\&+j_0\lambda^2\left(1-
    \int_{r}^{R} d\rho'
    \;\tilde{C}(\rho,\rho')\right), \label{eq:curr_int2}
\end{split}
\end{equation}
where $\tilde{C}(\rho,\rho')=-\int_{0}^{\infty} dk \frac{1}{k\rho'} J_0(k\rho) J_0(k\rho')
     \left(1 - \frac{1 - e^{-kh}}{kh}\right)$ and $\tilde{Q}(\rho,\rho')=\tilde{C}(\rho,\rho')\left(\rho'^2+\lambda^2\right)$.
Here, we average the integral (\ref{eq:expansion}) over the $z$-coordinate, because we assumed that the current density $j$ does not change significantly across the SC ring thickness and consider it as a function of only $\rho$-coordinate. Eq. (\ref{eq:curr_int2}) can be integrated numerically by sampling the function $j(\rho)$ on grid of $N$ equidistant points $\rho_i$ in the interval $(r,R)$. In this way the function $j(\rho)$ becomes a vector $j$ with $N$ components $j_i=j(\rho_i)$, and the integral kernels are transformed into the $N\times N$ matrices $\mathcal{Q}$ and $\mathcal{C}$ of the elements $\mathcal{Q}_{i,j}=\tilde{Q}(\rho_i,\rho'_j)$ and $\mathcal{C}_{i,j}=\tilde{C}(\rho_i,\rho'_j)$, respectively. Then, Eq.~(\ref{eq:curr_int2}) can be reduced to the matrix  equation for the components of the vector $j_i$ which has the following solution (for details see Refs. \cite{Brandt,Kharlan2024}):
\begin{equation}
j_i=j_0\lambda^2\sum_{j}\mathcal{P}_{i,j}\label{eq:J_final}
\end{equation}
The matrix $\mathcal{P}$ is defined as:
\begin{equation}
    \mathcal{P}=(
    \lambda^2\mathcal{I}-\Delta\rho\,\mathcal{Q})^{-1}(\mathcal{I}-\Delta\rho\,\mathcal{C})\label{eq:P},
\end{equation}
where $\Delta\rho=(R-r)/N$ and $\mathcal{I}$ denotes the identity matrix.
Finally, the distribution of the magnetic stray field produced by the eddy currents of the SC ring can be determined from the Biot--Savart law:
\begin{equation}
    \begin{split}
    B_{{\rm sc},\rho}(\rho,z)=-\frac{\mu_0}{4\pi}\int_{V_{\rm r}}&d^3\mathbf{r'}S(\rho,\rho',z,z')\\&\times j(\rho') \cos\varphi' (z-z'),\\
    B_{{\rm sc},z}(\rho,z)=\frac{\mu_0}{4\pi}\int_{V_{\rm r}}&d^3\mathbf{r'}S(\rho,\rho',z,z')\\&\times j(\rho') (\rho \cos\varphi'-\rho'),
    \end{split} \label{eq:stray_field}
\end{equation}
where the integration goes over the volume of the SC ring $V_{\rm r}$ with $d^3\mathbf{r'}=\rho'd\rho' d\varphi'dz'$ and $S(\rho,\rho',z,z')=(\rho^2+\rho'^2-2\rho\rho'\cos\varphi'+(z-z')^2)^{-\frac{3}{2}}$.

\begin{figure}
\includegraphics[width=0.95\columnwidth]{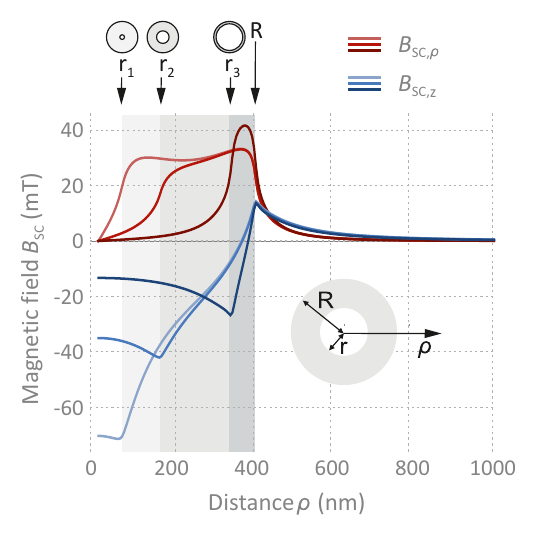}
\caption{Profiles of the magnetic field components $B_{{\rm sc},z}$ (blue lines) and $B_{{\rm sc},\rho}$ (red lines) generated by SC rings with the same external radius $R = 405$~nm and different internal radii $r$: $r_1 = 61$~nm, $r_2 = 160$~nm, and $r_3 = 344$~nm. The field profiles are plotted along the radial direction, starting from the center of the SC ring at $\rho = 0$. Gray vertical strips indicate the region occupied by the SC ring: $r < \rho < R$.} 
\label{fig:SC field}
\end{figure}

The exemplary profiles of the magnetic stray field components
$B_{\rm sc}$ produced by SC ring of the external radius $R=405$~nm and different internal radii $r$: $r_1 = 61$~nm, $r_2 = 160$~nm, and $r_3 = 344$~nm are presented in Fig.~\ref{fig:SC field}. Calculations were done for the  rectangular pulse of applied electric field  $Et_0=7\times10^{-9}$~V/m~s. 
We assumed the London penetration depth $\lambda = 50$ nm, what is  realistic for Nb \cite{Lambda}. Blue and orange solid lines corresponds to the $B_{{\rm sc},\rho}$ and $B_{{\rm sc},z}$ components, respectively, plotted along the radial direction. It is clear from Fig.~\ref{fig:SC field} that the radial component of the SC field  $B_{sc,\rho}$ promotes in-plane static magnetization configuration in FM under SC ring, while the normal component $B_{{\rm sc},z}$ makes favorable out-of-plane magnetization near the ring edges, which has opposite directions inside and outside of the ring. In other words, the magnetic stray field of SC ring has the same symmetry as Néel-type skyrmion magnetization, what allows us to expect that application of the magnetic field generated by SC ring to FM disk can stabilize the skyrmion texture. To check this hypotheses and establish the conditions for the skyrmion stabilization, we have formulated a theoretical model, which is described in Sec.~\ref{sec:Skrmion_theory}.  

\subsection{Skyrmion stabilization by the field of superconducting ring\label{sec:Skrmion_theory}}

Let us investigate the  energy of the ferromagnetic layer (or disk of the radius $R_{d}\gg R$) to find the magnetization configurations that can be stabilized in the FM layer by the magnetic field generated by the SC ring. We assume that the layer is thin, so that the magnetization can be considered uniform along the out-of-plane 
$z$-coordinate. The magnetic energy functional for the disk of the radius $R_d$ and thickness $d$ can be written as
\begin{equation}
  E[\mathbf{M}]=d\int_{S_{\rm d}}d^2\mathbf{r}\; \varepsilon\big(\mathbf{M}(\mathbf{r})\big), \label{eq:functional} 
\end{equation}
 where  the integration goes over the area  $S_{\rm d}$ of the disk's cross-section in the $\rho\!-\!\varphi$ plane with $d^2\mathbf{r}=\rho d\rho d\varphi$.  The energy density $\varepsilon$ consists of the exchange $\varepsilon_{\rm ex}$, magnetostatic $\varepsilon_{\rm ms}$, perpendicular magnetic anisotropy $\varepsilon_{\rm an}$ and Zeeman $\varepsilon_{z\rm }$ contributions. 

To derive the explicit expression for the magnetic energy density, it is convenient to represent the magnetization in spherical coordinates with the azimuthal and polar angles $\theta(\rho,\varphi)$ and $\phi(\rho,\varphi)$: 
$\mathbf{M} = M_{\rm s}(\sin\theta \cos\phi\,\hat{\mathbf{x}}+\sin\theta \sin\phi\,\hat{\mathbf{y}}+ \cos\theta\,\hat{\mathbf{z}})$, where $M_{\rm s}$ is the saturation magnetization. Then, the exchange energy density takes a form \cite{Sheka_2015}:
\begin{equation}
\varepsilon_{\rm ex}=A_{\rm ex}\big((\nabla\theta)^2+(\nabla\phi)^2\sin^2\theta\big), \label{eq:energy_ex}
\end{equation}
where $A_{\rm ex}$ is exchange stiffness. The magnetostatic energy density within the limit of the ultrathin disk is close to the energy density of the infinite film: $\varepsilon_{\rm ms}=\tfrac{1}{2}\mu_0M_{\rm s}^2\cos^2\theta$ -- we neglected here the contribution of volume magnetic charges. The energy density of uniaxial perpendicular magnetic anisotropy is $\varepsilon_{\rm an}=-K_{u\rm }\cos^2\theta$, where $K_{\rm u}$ is anisotropy constant. Since the magnetostatic interactions are responsible for the sample shape anisotropy, we can define the effective anisotropy energy density:
\begin{equation}
-K^{\rm eff}_{\rm u}\cos^2\theta=\varepsilon_a+\varepsilon_{\rm ms}=- \big(K_{\rm u}-\tfrac{1}{2}\mu_0 M_{\rm s}^2\big)\cos^2\theta. \label{eq:energy_an}
\end{equation}
The Zeeman energy is defined by applied magnetic field, which is here the magnetic stray field produced by the SC ring. Remembering that the SC stray field creates conditions favorable for stabilization of the Neel-like skyrmion texture (Fig. \ref{fig:SC field}), we can conclude that the Zeeman interaction is responsible for the skyrmion formation in the absence of DMI. Since the SC magnetic stray field has radial symmetry, the Zeeman energy density can be written in the following form:
\begin{equation}
\varepsilon_z=-M_{s}(B_\rho \sin \theta+B_z\cos\theta). \label{eq:energy_zeeman}
\end{equation}
To find the stable magnetization texture, we have to minimize the functional $E[\mathbf{M}]$ with respect to the $\theta$ and $\phi$ magnetization angles solving the Euler-Lagrange equations for these angles. Due to the symmetry of stray field  (\ref{eq:stray_field}),  we look for axially symmetric magnetization configurations, where $\mathbf{M}(\rho,\varphi)$ is a specific function. It means that the polar and azimuthal angles for magnetization vector are: $\theta(\rho,\varphi)$= $\theta(\rho)$ and $\phi(\rho,\varphi)=\varphi+\varphi_0$, where $\varphi_0=0$ or $\pi$ for N\'eel skyrmion. The magnetization texture $\mathbf{M}(\mathbf{r})$ which extremizes the functional (\ref{eq:functional}) is then expressed  by the scalar function $\theta(\rho)$ which depend on one coordinate only. It simplifies the variational problem for Eq.~\ref{eq:functional} to the solution of one-dimensional Euler–Lagrange equation for $\theta(\rho)$ function (see Ref.~\onlinecite{Guslienko_2018}) with boundary conditions: $\left.\theta(\rho)\right|_{\rho=0}=\pi$, $\left.\tfrac{d}{d\rho}\theta(\rho)\right|_{\rho=R_d}=0$ \cite{Du_2013}. 
Despite this simplification, analytical solution of the Euler–Lagrange equation is not possible. For this reason, we looked for a solution in the form of a trial function. We have used the so-called DeBonte ansatz \cite{DeBonte1973}:
\begin{equation}
\tan\big(\tfrac{1}{2}\theta(\rho)\big)=\frac{R_s}{\rho}\,\exp\big({-\tfrac{1}{\Delta}(\rho-R_{\rm s})}\big), \label{eq:DeBonde}
\end{equation}
which is a quite accurate description of the texture for the N\'eel type skyrmions with two parameters:  the width of skyrmion domain wall $\Delta$  and its radius  $R_{\rm s}$ (defined as a center of domain wall, where $\theta(\rho=R_{\rm s})=\pi/2$). In our studies, we fixed the width of the domain wall because it depends mostly on the competition between material parameters out-of-plane anisotropy and exchange stiffness \cite{cullity2009}: $\Delta=\sqrt{{A_{\rm ex}}/{K_{\rm u}^{\rm eff}}}$.
For the N\'eel skyrmion ansatz (\ref{eq:DeBonde}), the energy (\ref{eq:functional}) is a function of one parameter -- the  radius of skyrmion domain wall $R_{\rm s}$ (see Ref.~\onlinecite{Guslienko_2018}, \onlinecite{Navas2019}):
\begin{equation}
\begin{split}
    E(R_{\rm s})& = 2\pi d  \int_0^{R_d} \rho \, d\rho \Bigg[\frac{2 A_{\rm ex}}{M_{\rm s}^2} \left( \frac{1}{\rho^2} + \frac{1}{\rho\Delta}  + \frac{1}{\Delta^2} \right) M_{\rho}^2 \\
     &  -  \left( B_{{\rm sc},\rho} M_{\rho} + B_{\rm{sc},z} M_z \right) \Bigg]-\pi d K_{\rm u}^{\rm eff} R_{\rm d}^2,
\label{eq:energy_final}
\end{split}
\end{equation}
where $M_\rho(\rho)=M_{\rm s}\sin(\theta(\rho)) 
$ and $M_z(\rho)=M_{\rm s}\cos(\theta(\rho)) 
$ are the components of magnetization vector in cylindrical coordinate system: $\mathbf{M}=M_\rho\hat{\mathbf \rho}+M_z\hat{\mathbf z}$. The components of the magnetic stray field (\ref{eq:stray_field}): $B_{\rm{sc},\rho}(\rho)$  and $B_{\rm{sc},z}(\rho)$ are taken in the middle plane of thin FM layer.

Using the standard procedure (i.e., looking for the solutions of the equation $dE/dR_{s}=0$ satisfying to the condition  $d^2E/dR_{s}^2>0$), we can minimize the energy (\ref{eq:energy_final}) with respect to the  radius of skyrmion domain wall $R_{\rm s}$ to find the radius $R_{\rm s}=R_{\rm s,0}$ of the stable skyrmion. We consider that the skyrmion can be generated in the system if the energy (\ref{eq:energy_final}) has a minimum for some finite value of $R_{\rm s}$ inside the FM disk: $0<R_{\rm s}<R_{d}$.

\subsubsection*{Micromagnetic simulations}

The theoretical studies of skyrmion stabilization  are cross-checked by micromagnetic simulations. We use our own version of Mumax3~\cite{TheMuMax3,Leliaert2014ASimulations} called AMUmax~\cite{MathieuMoalicAMUmax}
, which solves the Landau--Lifshitz--Gilbert equation:
\begin{equation}
 \frac{\text{d}\mathbf{m}}{\mathrm{d}t}= 
 \frac{|\gamma| \mu_0}{1+\alpha^{2}} \left(\mathbf{m} \times \mathbf{H}_{\mathrm{eff}} + 
 \alpha  \mathbf{m} \times 
 (\mathbf{m} \times \mathbf{H}_{\mathrm{eff}}) \right),
\end{equation}
where $\textbf{m} = \textbf{M} / M_{\mathrm{s}}$ is the normalized magnetization, $\textbf{\text{H}}_{\mathrm{eff}}$ is the effective magnetic field acting on the magnetization, $\gamma$ is the gyromagnetic ratio, $\mu_0$ denotes the vacuum permeability and $\alpha$ is the Gilbert damping.
The following components contribute to the effective magnetic field $\textbf{H}_{\mathrm{eff}}$: demagnetizing field in magnetostatic approximation $\textbf{\text{H}}_{\mathrm{ms}}$, exchange field $\textbf{\text{H}}_{\mathrm{ex}}$, uniaxial magnetic anisotropy field $\textbf{\text{H}}_{\mathrm{an}}$, and external magnetic field $\textbf{\text{H}}_{\mathrm{sc}}=1/\mu_0\textbf{\text{B}}_{\mathrm{sc}}$ produced by SC ring. 
\begin{equation}
  \textbf{H}_{\mathrm{eff}} =
  \textbf{H}_{\mathrm{ms}} + \textbf{H}_{\mathrm{ex}} + \textbf{H}_{\mathrm{an}} + \textbf{H}_{\mathrm{sc}}.
\end{equation}
The exchange and anisotropy fields are defined as
\begin{equation}
  \textbf{H}_{\mathrm{ex}} = 
  \frac{2A_{\mathrm{ex}}}{\mu_0 M_{\mathrm{s}}} \Delta \textbf{m},\;
  \textbf{H}_{\mathrm{an}} =\frac{2K_{\mathrm{u}}}{\mu_0 M_{\mathrm{s}}} m_z \hat{\textbf{z}}
\label{Eq:Fields}
\end{equation}
and magnetostatic field is computed numerically from general formula: 
\begin{equation}
\textbf{H}_{\rm ms}(\textbf{r})=-M_{\rm{s}}\int\textbf{N}(\textbf{r}-\textbf{r}')\cdot\textbf{m}(\textbf{r}')d^3\textbf{r}',
\end{equation} 
where the demagnetizing tensor is:
\begin{equation}
    \textbf{N}(\textbf{r}-\textbf{r}')=\frac{1}{4\pi}\nabla\nabla'\frac{1}{|\textbf{r}-\textbf{r}'|}
\end{equation}
The external field $\mathbf{H}_{\rm sc}=1/\mu_0 \mathbf{B}_{\rm sc}$ is calculated as it is presented in Sec.~\ref{sec:SC_theory} and introduced to micromagnetic solver. We neglect the thermal effects.

To calculate the skyrmion magnetic energy, we use the frozen spin technique with the \textit{frozen spin function} built into Mumax3 and AMUmax. In the sequence of calculations, we select the different trial  radii of skyrmion domain wall $R_{\rm s}$ and frozen the magnetization  ($\theta=\pi/2, \,\phi=\varphi$) in very narrow ring (one cuboid of the mesh) of the radius $R_{\rm s}$.
By varying the radius of the artificial ring, we compute the dependence energy of the skyrmion as a function of its radius $E(R_{\rm s})$. This allows to find the  the radius for which the energy is minimized. Then we remove the constraint of frozen magnetization and allow the whole texture to relax freely. As a result, we are able to find a stable N\'eel skyrmion  magneization configuration in the magnetic field produced by SC ring.

The material parameters and the dimensions of considered structure, used in the semi-analytic calculations and micromagnetic simulations are given in Sec.~\ref{sec:Results}.

\section{Results\label{sec:Results}}

\begin{figure*}
\centering \includegraphics[width=0.95\linewidth]{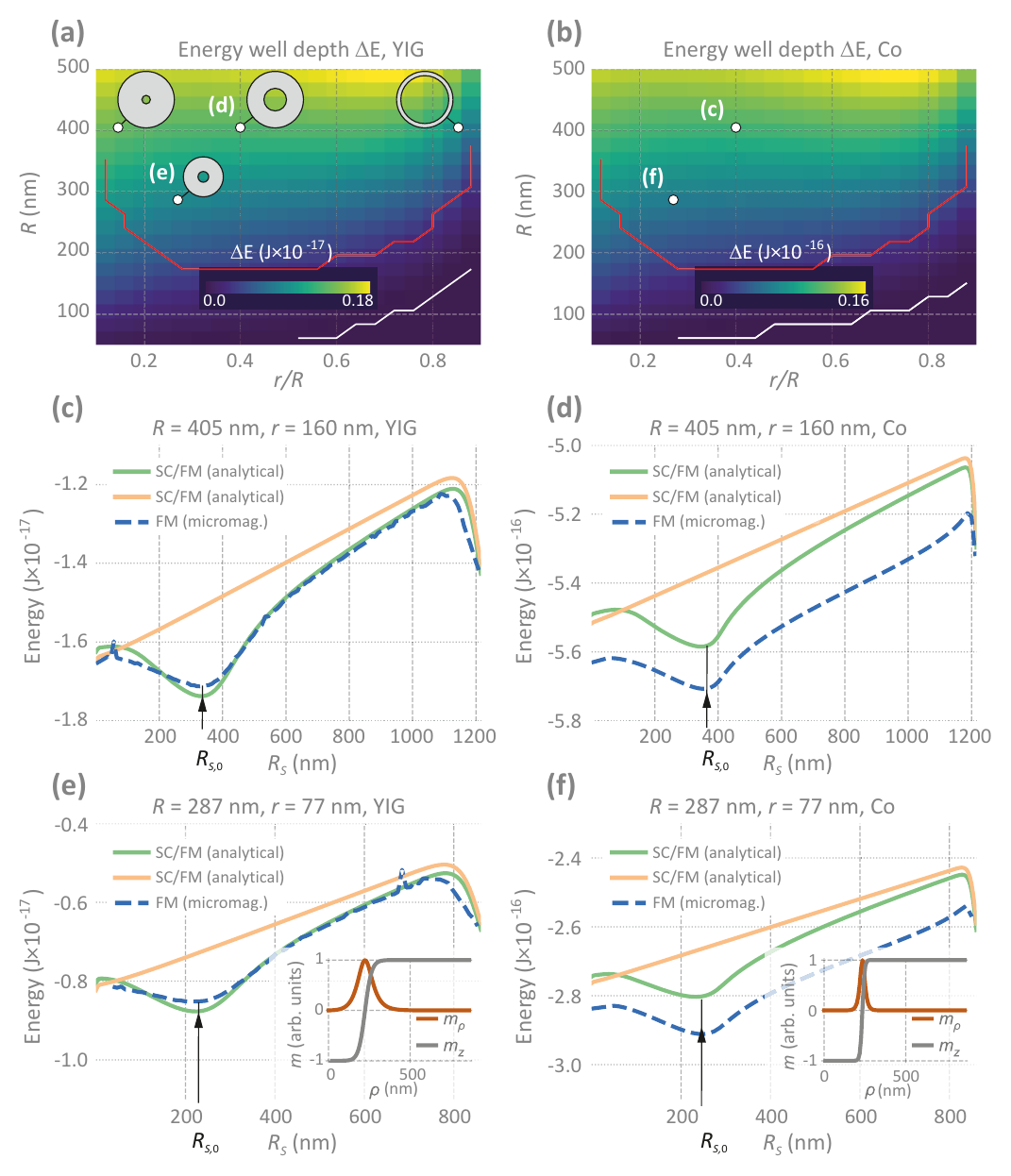}
\caption{Dependencies of the energy well depth for the skyrmion texture, $\Delta E$, on the geometry of the superconducting (SC) ring: the outer ring radius $R$ and the ratio between the inner and outer radii $r/R$, calculated for Ga:YIG (a) and Co (b). The uniformly colored dark blue area below the white line denotes the region where the energy well is absent ($\Delta E = 0$), and the skyrmion cannot be stabilized. In the region between the red and white lines, the energy well has a finite depth, but the current exceeds the critical value for Nd, given the electric field pulse profile $Et_0 = 11.25 \times 10^{-9}$~Vs/m. In the region above the red line, the skyrmion can be stabilized by the superconducting current. For both Ga:YIG (c,e) and Co (d,f), we show the dependence of the skyrmion energy on the assumed radius of skyrmion domain wall $R_{\rm s}$, for two selected SC ring sizes (indicated by white points in (a) and (b)). The energy well is deeper for larger outer radius $R$ and for narrower ring width (i.e., $r/R \approx 0.7$) -- compare (c,d) to (e,f). The absence of a well in the absence of the SC ring’s stray field demonstrates the necessity of the SC ring for skyrmion stabilization. Green and dashed blue lines represent the results of semi-analytical and micromagnetic calculations, respectively. Insets in (c) and (d) show the profile of the stable skyrmion texture, with $R_{\rm s}$ corresponding to the energy minimum: $R_{\rm s}=R_{\rm s,0}$.} 
\label{fig:Skyrmion everything}
\end{figure*}
In our calculations we have used a SC ring with thickness $h=100$ nm and varying outer radius $R=50-500$ nm and inner radius $r=0.1R-0.9R$. The SC ring is made of Nb with the London penetration depth $\lambda=50$ nm \cite{Lambda}. For the FM disk, we explored two commonly used materials with significantly different parameters: Ga:YIG (saturation magnetization $M_{\rm s}=160$~kA/m, exchange stiffness $A_{\rm ex}=1.37$~pJ/m, PMA $K_{\rm u}=756$ J/m$^3$, disk thickness $d=6$ nm \cite{Bottcher_2022,YIG}) and Co ($M_{\rm s}=1.4$~MA/m, $A_{\rm ex}=30$~pJ/m, $K_{\rm u}=1.39$~MJ/m$^3$, $d=0.75$ nm \cite{Co}). To eliminate the influence of the disk edge on the skyrmion, we set the radius of the disk to $R_{\rm d}=3R$. Calculations, whose results are presented in Fig.~\ref{fig:Skyrmion everything}, were done for the fixed profile of the electric field pulse $Et_0=11.25\cdot10^{-9}$~V~s/m.

For the skyrmion described by the DeBonte ansatz, we calculated the energy of its texture assuming a fixed domain wall width: $\Delta = \sqrt{{A_{\rm ex}}/{K_{\rm u}^{\rm eff}}}$, and examined how it depends on the assumed radius of the skyrmion domain wall: $E(R_{\rm s})$ -- see Eq. (\ref{eq:energy_final} ) and Fig.\ref{fig:Skyrmion everything}(c–f). We then searched for the minimum of the $E(R_{\rm s})$ dependence and identified its position $R_{\rm s,0}$ and depth $\Delta E$. The parameter $R_{\rm s,0}$ is considered as a radius of a stable skyrmion and $\Delta E$ a measure of its stability against thermal or microwave excitation, according to the Arrhenius law. It is worth noting that at the boundaries of the domain, $0 < R_{\rm s} < R_{\rm d}$, we also observe the energy wells. 
However, when $R_{\rm s} = 0$ or $R_{\rm s} = R_{\rm d}$, the FM disk of radius $R_{\rm d}$ is  in single-domain out-of-plane magnetization configuration, and the skyrmion texture vanishes. The energy minimum $E(R_{\rm s,0})$, corresponding to the stable skyrmion, is separated by two maxima: a lower one (for $R_{\rm s}$ close to 0) and a higher one (for $R_{\rm s}$ close to $R_{\rm d}$). Therefore, the depth of the energy well, $\Delta E$ is determined with respect to the lower maximum — see Fig.~\ref{fig:Skyrmion everything}(c). The results of the semi-analytic calculations (solid green lines in Fig.~\ref{fig:Skyrmion everything}(c–f)) were compared to the outcomes of micromagnetic simulations (dashed blue lines in Fig.~\ref{fig:Skyrmion everything}(c–f)). The values of $R_{\rm s,0}$ and $\Delta E$ are in good agreement, confirming that the presented semi-analytic method can be reliably used to investigate the size and stability of magnetic skyrmions in considered SC/FM hybrid nanostructures. However, the analytical approach tends to overestimate the magnetic energy of the skyrmion textures (i.e., it does not minimize the energy for a given $R_{\rm s}$), due to the assumptions and constraints introduced (due to the fixed width of the domain wall).

First, we aim to investigate how the size of the SC ring influences  skyrmion stability. The dependence of the depth of the energy well $\Delta E$ on the outer ring radius $R$ and the ratio between the inner and  outer radii $r/R$ were calculated using semi-analytic model for Ga:YIG (Fig.~\ref{fig:Skyrmion everything}a) and Co (Fig.~\ref{fig:Skyrmion everything}b). Assuming that the SC ring is excited the same way, regardless of its size, we fixed the height and duration of the electric field pulse that induces the SC current: $Et_0 = 11.25 \times 10^{-9}$~V s/m.
In dark blue area, below the white line,  $\Delta E = 0$ and the skyrmion can not be stabilized. Area of non-zero $\Delta E$ (i.e., the region above the white line) can be extended by increasing of the electric field pulse value, what leads to the enhancement of the SC stray magnetic field. Wherein, one should remember, that induced current can not exceed the critical value, above which the Meissner state is destroyed in the SC ring. 
 For the taken profile of the electric field pulse $E t_0$, the region, where, the maximum current density is smaller than the critical value for Nb is delimited by the red line. This area (i.e. the region above the red line) will increase with the decrease of $Et_0$. It is evident that, in the case of a small and narrow SC ring (bottom-right corner of the plots in Fig.~\ref{fig:Skyrmion everything}(a,b), stabilizing the skyrmion becomes more challenging, with the primary limiting factor being the  critical current density of the superconducting material.
 
We found also that the radius of the skyrmion domain wall $R_{\rm s0}$ is mostly determined by the outer size of SC ring $R$, and its relative width ($(R-r)/R=1-r/R$) plays minor role (see Supplemental Material). This finding is easy to understand if one looks at Fig.~\ref{fig:SC field}: maximum the of radial component of the SC stray field $B_{\rho}$, which corresponds to the middle of domain wall (in-plane magnetization configuration), is always close to $R$.

It is clear from Fig.~\ref{fig:Skyrmion everything}(a,b) that, for a fixed ratio $r/R$, a larger outer ring radius $R$ results in a more stable skyrmion. In such a case, the skyrmion is simply larger (higher values of $R_{\rm s0}$). Therefore, any changes in the magnetic configuration of the stabilized skyrmion domain wall will require more energy and  the depth energy minium $\Delta E$ must be larger. Moreover, the skyrmion stability increases as the ring becomes thinner (for fixed $R$), up to a certain value of $r/R$ (about 0.7 for YIG and 0.75 for Co). However, for very thin SC rings, we see the essential decrease of the skyrmion stability. This effect can be attributed to the manner in which the SC ring generates the stray magnetic field—see Fig.~\ref{fig:SC field}. For narrow rings (i.e., larger $r/R$ ratios), the radial component of the stray field, $B_{\rm sc,\rho}$, is enhanced in the vicinity of the domain wall at $R_{\rm s,0} \approx R$, which slightly improves skyrmion stabilization. Conversely, for the SC wider rings (i.e., smaller $r/R$ ratios), a nonzero $B_{\rm sc,\rho}$ persists within the skyrmion core, which tends to destabilize the texture.
The decrease of $\Delta E$ for $r/R\approx 1$
is dictated by the reduction of the SC current in the ring when its width $R(1-r/R)$ is comparable to the London penetration depth $\lambda$. We note that width of the SC ring is the smallest in the bottom-right corner of the plots in Fig.~\ref{fig:Skyrmion everything}(a,b), below the white line. In this region the stray magnetic field is too weak to induce any minimum in the $E(R_{\rm s})$ dependence.

Let us now discuss in more detail the dependence of the energy of skyrmion texture on the assumed size of skyrmion, $E(R_{\rm s})$, as presented in Fig.~\ref{fig:Skyrmion everything}(c--f). The energy profiles $E(R_{\rm s})$ were plotted for selected superconducting ring geometries: $R = 405\,\text{nm}$, $r = 160\,\text{nm}$ (with $r/R = 0.4$) — see Fig.~\ref{fig:Skyrmion everything}(c,d), and $R = 287\,\text{nm}$, $r = 77\,\text{nm}$ (with $r/R = 0.27$) — see Fig.~\ref{fig:Skyrmion everything}(e,f). Two materials were considered: Ga:YIG — Fig.~\ref{fig:Skyrmion everything}(c,e), and Co — Fig.~\ref{fig:Skyrmion everything}(d,f).
The orange lines show, for reference, the energy of a skyrmion texture in a  FM disk in the absence of the external magnetic field generated by the SC ring. In this case, no energy minimum appears for $0 < R_{\rm s} < R_{\rm d}$, indicating that the skyrmion cannot be stabilized. The positive slope of $E(R_{\rm s})$ implies that, once imposed, the skyrmion texture gradually annihilates by shrinking its radius to zero. This linear dependence of $E(R_{\rm s})$ arises from the fact that, without an external field, the energy is determined solely by the domain wall, and is therefore proportional to its circumference: $2\pi R_{\rm s}$. 
However, placing the FM disk in the stray magnetic field generated by the SC ring results in the appearance of an energy minimum (dashed blue and solid green lines), which enables the skyrmion stabilization at the radius corresponding to this minimum. For both YIG and Co, increasing the SC ring size $R$ and narrowing its relative width, $(R - r)/R = 1 - r/R$, deepens the energy well — i.e., enhances the skyrmion stabilization — as discussed above. This demonstrates a universal mechanism for controlling skyrmion stability via the geometry of the SC ring and the SC current, which can be modified by electric field pulses.

The insets in Fig.\ref{fig:Skyrmion everything} (c,d) present normalized magnetization profiles of the skyrmion texture at the energy minimum: $R_{\rm s}=R_{\rm s,0}$. Red and gray lines show $m_\rho$ and $m_z$ components, respectively, plotted using Eq. (\ref{eq:DeBonde}). Wherein, $\rho=R_{s}$ corresponds to zero value of $m_z$, i.e. the center of the radial domain wall. As one can see, the width of domain wall, the depth of energy well  and, therefore, the absolute stability of the skyrmion, expressed by the value of $\Delta E$, are significantly different for systems based on YIG and Co ferromagnetic materials, even if the have the same geometrical parameters. Thus, we are smoothly approaching to the question of how the material parameters of a ferromagnet affect skyrmion stabilization and which materials would be optimal for  applications. 

There are three main material parameters whose interplay determines the suitability of a material for stabilizing skyrmions using the stray magnetic field of the  SC nanostructure: saturation magnetization ($M_{\rm s}$), exchange stiffness ($A_{\rm ex}$), and uniaxial magnetic anisotropy ($K_{\rm u}$) (see Supplemental Material for details). Note again that the considered materials do not exhibit any DMI. Our analysis has shown that larger values of $M_{\rm s}$ and smaller values of $A_{\rm ex}$ increase the energy difference $\Delta E$, and are therefore favorable for the skyrmion stabilization. This behavior can be understood by considering that the skyrmion is stabilized via Zeeman interaction, where the stray magnetic field of the SC ring  affects the magnetization. As a result, increasing $M_{\rm s}$ enhances the strength of this interaction and promotes the skyrmion stabilization. The requirement for small $A_{\rm ex}$ stems from the need to reduce the stiffness of the magnetic texture, which has to be reversed in the center of the skyrmion relative to its outer regions. The influence of the out-of-plane magnetic anisotropy is such that lower anisotropy values facilitate the skyrmion stabilization, as they allow the magnetization to rotate more easily. However, the uniaxial anisotropy constant must still exceed the shape anisotropy energy in order to maintain an out-of-plane magnetization configuration far from the skyrmion domain wall.

The choice of the investigated FM materials pursued the goal of showing these features. Ga:YIG is characterized by relatively small values of both $M_s$ and $A_{ex}$. Wherein, this material has low damping of spin waves, what makes it attractive for using in real magnonic devices. Cobalt is characterized by both relatively high values of $M_{\rm s}$ and $A_{\rm ex}$. It is one of the most popular and investigated FM materials with well established production technology. 
For considered parameters Co provides significantly larger stability of the skyrmion (larger $\Delta E$) and broader range of suitable geometrical parameters of an SC ring (Fig.\ref{fig:Skyrmion everything}(a,b)) due to the much larger value of $M_{\rm s}$ in comparison with Ga:YIG, despite the smaller $A_{ex}$ and lower $K_{\rm u}$ of Ga:YIG. Thus, in this case, the Zeeman interaction plays a crucial role in the determination of the skyrmion stability.  

\begin{figure}[t]
\includegraphics[width=0.95\columnwidth]{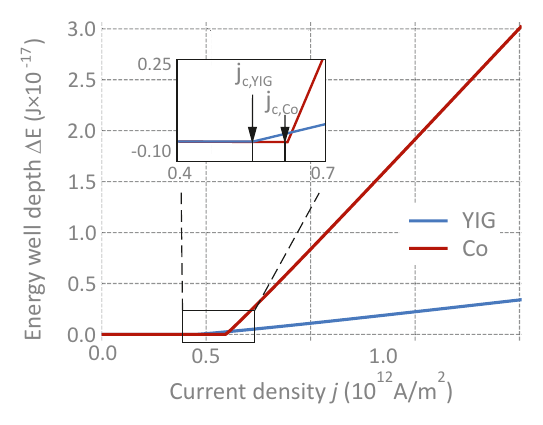}
\caption{Dependence of the energy well depth for the skyrmion texture, $\Delta E$, on the maximum current density in the SC ring of fixed size (outer radius: $R~=~405$~nm, inner radius: $r~=~160$~nm), calculated for Ga:YIG (blue line) and Co (red line). The inset shows a zoomed-in region near the threshold current value, where the stray field generated by the SC ring becomes sufficient to stabilize the skyrmion.}
\label{fig:current}
\end{figure}

Finally, let us discuss the most notable advantage of skyrmion stabilization by the stray magnetic field produced by the superconducting ring: the possibility to control skyrmion stability via the current flowing through the ring, which can be tuned by pulses of the applied electric field. In particular, an increase in current enhances the SC stray magnetic field and, consequently, the Zeeman energy, which in turn deepens the energy potential well. The dependence of the energy barrier $\Delta E$ on the maximum current density in the ring with $R = 405$ nm and $r = 160$ nm is presented in Fig.~\ref{fig:current} for Ga:YIG (blue line) and Co (red line). The inset shows a zoomed-in region near the current densities at which $\Delta E$ becomes non-zero, indicating the critical value $j_{\rm c}$, above which the skyrmion can be stabilized in the system. The value of $j_{\rm c}$, as well as the rate of increase of $\Delta E$ with the current density $j$ increasing, is determined by the interplay between the FM parameters $M_{\rm s}$ and $A_{\rm ex}$, as discussed above.
Interestingly, the $\Delta E(j)$ curves for Ga:YIG and Co intersect. This behavior arises from the fact that, at low currents, the SC stray magnetic field and the resulting Zeeman interaction are weak, so the stability is primarily governed by the exchange interaction. However, at higher current densities, the contribution of the Zeeman energy becomes more significant, strongly affecting skyrmion stability. This explains why, at low current values, the skyrmion is more stable in Ga:YIG due to its low exchange stiffness $A_{\rm ex}$, while at higher current densities, Co becomes more favorable for the skyrmion stabilization because of its large saturation magnetization $M_{\rm s}$.

\section{Conclusions}
The stabilization of Néel-type skyrmions in a hybrid system consisting of  ferromagnetic disk without Dzyaloshinskii–Moriya interaction, placed under a superconducting ring, has been studied both  analytically and numerically. It has been demonstrated that the application of an electric field pulse to the superconducting ring in the Meissner state generates a persistent current, which, in turn, induces a non-uniform magnetic field in the ferromagnetic disk. This field exhibits a spatial symmetry compatible with that of a Néel-type magnetic skyrmion, thus facilitating its stabilization. 

To establish criteria for selecting magnetic materials optimal for skyrmion formation in hybrid superconductor/ferromagnet structures, we investigated how key material parameters — saturation magnetization, out-of-plane magnetic anisotropy, and exchange stiffness — affect the skyrmion stabilization. We found that higher saturation magnetization enhances the Zeeman interaction between the superconductor ring and ferromagnet, promoting the skyrmion formation, while lower exchange stiffness facilitates the necessary magnetization reversal at the skyrmion core. Reduced out-of-plane magnetic anisotropy favors easier magnetization rotation, but must remain strong enough to overcome shape anisotropy and maintain out-of-plane magnetization alignment. We studied two typical ferromagnetic materials, Ga:YIG (low saturation magnetization and exchange stiffness) and Co (high values of both), and found that the skyrmion stability is significantly higher in Co, highlighting the key role of the Zeeman interaction and saturation magnetization. 

Finally, the most significant advantage of skyrmion stabilization by the superconducting ring's stray magnetic field is the ability to control the skyrmion stability through the current flowing in the ring, which can be adjusted by electric field pulses. In particular, an increase in the current density enhances the SC stray magnetic field, thereby enhancing the Zeeman energy and stabilizing the skyrmion. Wherein, there is a critical current density above which the skyrmion can be stabilized, offering a pathway for energy-efficient information writing and erasing through skyrmion nucleation and annihilation with a specified polarity, thus advancing the development of magnonic neuromorphic systems.

\begin{acknowledgments}
The work was supported from the grants of the National Science Center, Poland, No.  UMO-2021/43/I/ST3/00550.  K.G. acknowledges support by IKERBASQUE (the Basque Foundation for Science). The research of K.G. was funded in part by the National Science Center of Poland, project no UMO-2023/49/B/ST3/02920, the Spanish Ministry of Science, Innovation and Universities grant PID2022-137567NB-C21 /AEI/10.13039/501100011033 and by the Basque Country government under the scheme “Ayuda a Grupos Consolidados” (Ref. IT1670-22).  The authors thank Dr. Szymon Mieszczak for his advice and help in the numerical calculations of the stray field produced by the superconductor. 
\end{acknowledgments}

\vspace{0.5 cm}
\section*{Data availability}
Data supporting this study are openly available from the repository at https://doi.org/10.5281/zenodo.15358284 .


%

\clearpage


\section*{Supplementary material}

# Supplemental Material:
# Stabilization the of skyrmion in a hybrid magnetic–superconducting nanostucture

Julia Kharlan[1,2], Mateusz Zelent[1], Konstantin Guslienko[3], Vladimir O. Golub[2], Jarosław W. Kłos[1*]

[1]*ISQI, Faculty of Physics, Adam Mickiewicz University, Poznań, Poland*
[2]*V.G. Barykhtar Institute NASU of Ukraine, Kyiv, Ukraine   and*
[3]*Depto. Polimeros y Materiales Avanzados: Fisica, Quimica y Tecnologia,*
*Universidad del País Vasco, UPV/EHU, 20018 San Sebastian, Spain*
*EHU Quantum Center, University of the Basque Country, UPV/EHU, 48940 Leioa, Spain*
*IKERBASQUE, the Basque Foundation for Science, 48013 Bilbao, Spain*

## S1: IMPACT OF THE SC RINGSIZES AND THE SC CURRENT ON THE SKYRMION SIZE

The radius of the skyrmion domain wall $R_{s,0}$, stabilized by the stray field generated by the SC ring, is primarily determined by the outer ring radius $R$, while variations in the inner radius $r$ have only a minor effect on $R_s$. For fixed $R$, a decrease in $r$ leads to a slight reduction in $R_{s,0}$. This trend is clearly illustrated in Fig. S1(a), which shows the dependence of $R_{s,0}$ on the outer ring radius $R$ and the ratio $r/R$. The calculations were performed for a Ga:YIG FM disk and an electric field pulse of $Et = 11.25 \times 10^{-9}$ A s/m. This behavior can be attributed to profiles of the in-plane component of SC stray field, $B_\varphi$ – see Fig. 2 and related the discussion the main text.

The influence of the current magnitude is more pronounced. At low current densities, the SC stray field is too weak to nucleate a skyrmion. As the current increases, a critical threshold is reached at which the stray field becomes strong enough to create an energy well capable of stabilizing a skyrmion. This transition is abrupt: the skyrmion appears with a finite radius of domain wall $R_{s,0}$ (approximately equal to $R$), which remains nearly constant with further increases in current. This behavior is depicted in Fig. S1(b), where $R_{s,0}$ is plotted as a function of the maximum current density for rings with various geometries. Solid and dashed curves correspond to $r/R = 0.4$ and $r/R = 0.27$, respectively. These results further confirm that $R_s$ is largely insensitive to variations in the inner ring radius.

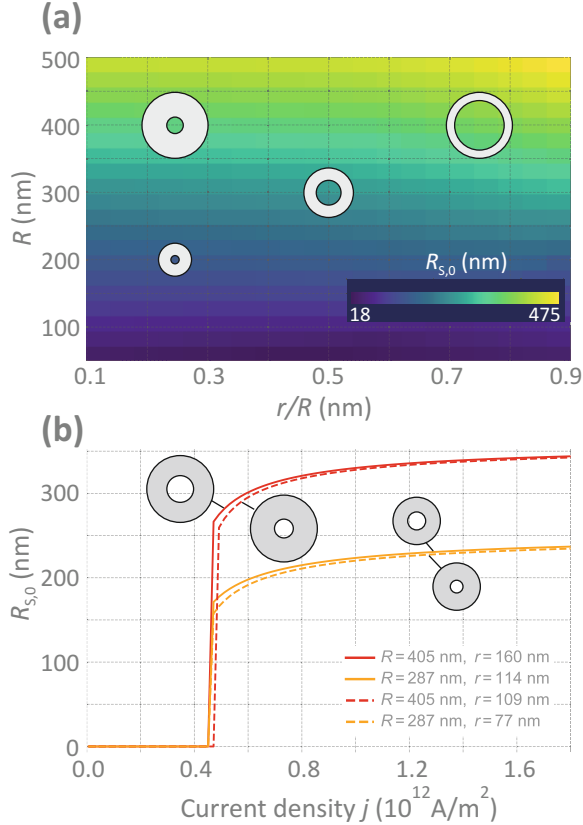

FIG. S1. (a) Dependence of the skyrmion domain wall radius, $R_s$, on the outer ring radius $R$ and the ratio of inner to outer radii, $r/R$, calculated for an electric field pulse of $Et = 11.25 \times 10^{-9}$ A·s/m in a Ga:YIG ferromagnetic (FM) disk. (b) Skyrmion radius as a function of the maximum current density in the ring for various values of $R$ and $r$. Solid and dashed lines correspond to $r/R = 0.4$ and $r/R = 0.27$, respectively. Red and orange curves represent $R = 405$ nm and $R = 287$ nm.

## S2: INFLUENCE OF THE MATERIAL PARAMETERS OF FM DISK ON THE SKYRMION STABILITY

Skyrmion stability, quantified by the depth of the magnetic energy well $\Delta E$, is strongly influenced by the material parameters of the ferromagnetic (FM) disk—specifically, the saturation magnetization $M_s$, perpendicular magnetic anisotropy $K_u$, and exchange stiffness $A_{ex}$. To isolate the impact of $M_s$, we computed the dependence of $\Delta E$ on the current density in the superconducting ring while keeping $K_u$ and $A_{ex}$ fixed at values corresponding to Ga:YIG, and systematically varying $M_s$. As shown in Fig. S2(a), an increase in $M_s$ leads to

---

[*] yuliia.kharlan@amu.edu.pl

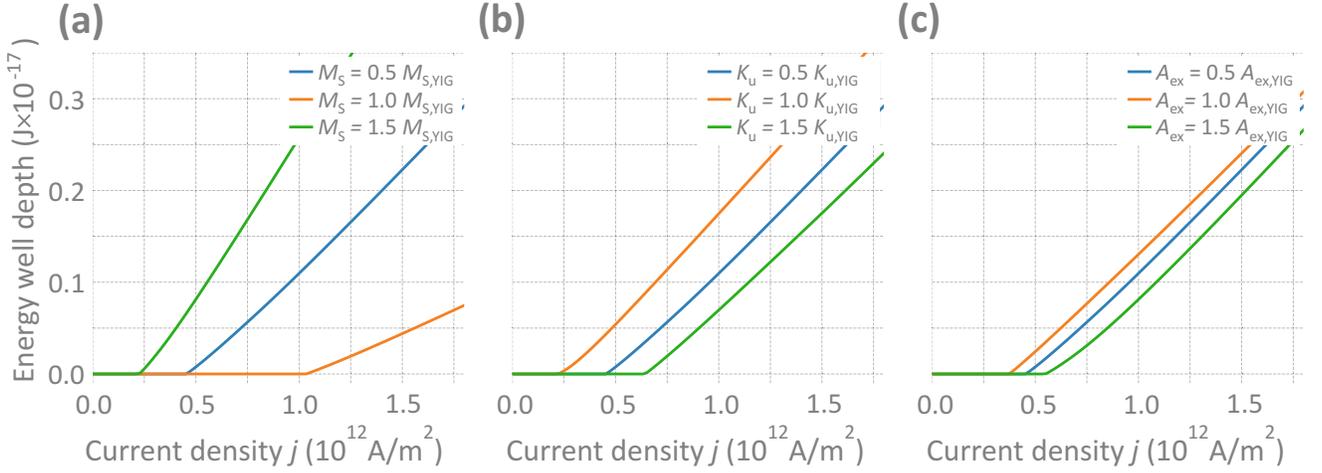

FIG. S2. The dependencies of the energy well deepness $\Delta E$ on the maximal current density $j$ in an SC ring of radii $R = 405$ nm and $r = 160$ nm. Subplots show how $\Delta E$ modifies with variation of the saturation magnetization (a), uniaxial magnetic anisotropy (b) and exchange stiffness (c). The blue lines correspond to $\Delta E$ of FM layer made of Ga:YIG. The orange and green lines corresponds to the varied value of $M_s$ ($K_u$ or $A_{ex}$), while the other parameters remind constant.

a reduction in the critical current required for skyrmion formation, along with a steeper increase in stability as current grows. This behavior arises from the enhanced Zeeman interaction between the SC stray field and the magnetization. In our system, the Zeeman energy is the dominant mechanism responsible for the continuous rotation of the magnetization by 180° from the disk center to its edge, ultimately stabilizing the skyrmion.

A similar analysis for the anisotropy constant, shown in Fig. S2(b), indicates that lower values of $K_u$ are more conducive to skyrmion formation. This is expected, as a reduced anisotropy decreases the energy barrier between up and down magnetization states, making it easier for the SC stray field to reorient the local magnetization. However, it is important to note that $K_u$ must remain above the threshold set by shape anisotropy in order to sustain out-of-plane magnetization both at the skyrmion core and its boundary. A similar trend is observed for the exchange stiffness: smaller values of $A_{ex}$ increase the flexibility of the magnetization texture, thus facilitating reorientation under the influence of the SC stray field.



\end{document}